\pdfoutput=1
\documentclass[%
 reprint,
 amsmath,amssymb,
 aps,
 superscriptaddress,
 preprintnumbers,nofootinbib,
]{revtex4-2}

\usepackage{graphicx}
\usepackage{dcolumn}
\usepackage{booktabs}
\usepackage{mathtools}
\usepackage{multirow}
\usepackage{amssymb}
\usepackage{mathrsfs}
\usepackage{bm}
\usepackage[caption=false]{subfig}
\usepackage[space]{grffile}
\def\cl{{\cal L}}

\def\[{\lfloor{\hskip 0.35pt}\!\!\!\lceil}
\def\]{\rfloor{\hskip 0.35pt}\!\!\!\rceil}

\begin{document}

\title{More on gravitational waves from double monodromy inflation}

\author{Medeu Abishev}
\email{abishevme@gmail.com}
\affiliation{Department of Theoretical and Nuclear Physics, Al-Farabi Kazakh National University, 71 Al-Farabi Ave., Almaty 050040, Kazakhstan}
\affiliation{Institute of Nuclear Physics, 1 Ibraginov Str., Almaty 050032, Kazakhstan}

\author{Aigerim Abylayeva}
\affiliation{Department of Theoretical and Nuclear Physics, Al-Farabi Kazakh National University, 71 Al-Farabi Ave., Almaty 050040, Kazakhstan}
\affiliation{Institute of Nuclear Physics, 1 Ibraginov Str., Almaty 050032, Kazakhstan}

\author{Andrea Addazi}
\email{addazi@scu.edu.cn}
\affiliation{Center for Theoretical Physics, College of Physics, Science and Technology, Sichuan University, 610065 Chengdu, China}
\affiliation{Laboratori Nazionali di Frascati INFN, Frascati (Rome), Italy}

\author{Yermek Aldabergenov}
\email{yermek.a@chula.ac.th}
\affiliation{Department of Theoretical and Nuclear Physics, Al-Farabi Kazakh National University, 71 Al-Farabi Ave., Almaty 050040, Kazakhstan}
\affiliation{Institute of Nuclear Physics, 1 Ibraginov Str., Almaty 050032, Kazakhstan}
\affiliation{Department of Physics, Faculty of Science, Chulalongkorn University, Phayathai Road, Pathumwan, Bangkok 10330, Thailand}

\author{Daulet Berkimbayev}
\affiliation{Department of Theoretical and Nuclear Physics, Al-Farabi Kazakh National University, 71 Al-Farabi Ave., Almaty 050040, Kazakhstan}

\date{\today}

\begin{abstract}

We further analyze phenomenological implications of double axion monodromy inflation proposed in Ref. Phys. Rev. D 104, L081302 (2021), in gravitational wave physics. We show that in addition to chiral gravitational waves (GW) originating from gauge field instability, the model also predicts significant amount of non-chiral, scalar-induced gravitational waves, both peaking at around the same frequencies. We find that although chiral GW density has much larger peak, non-chiral GWs can dominate away from the peak as they decay at a slower rate. This provides an interesting GW signature to be probed by future space-based interferometers such as LISA and DECIGO.

\end{abstract}

\maketitle

\section{Introduction}

Top-down (TD) approaches towards the UV completion of quantum gravity (QG) cross several obstacles which at the moment appear to be impassable. This highly motivates the exploration of bottom-up (BU) phenomenological QG effective models testable in cosmology and multi-messenger astrophysics \cite{Addazi:2021xuf}. Indeed, BUQG can provide an important guidance for TDQG, while TDGQ inspiring new BUQG effective models.  

Moreover, the inflation dynamics as well as inflaton(s) origin remain unknown, representing important questions for any UV QG approaches. Perhaps, the most dramatic situation is in the context of String Theory (ST) which provides a multitude of possible inflation scenarios. At the moment, it is impossible to select one ST prediction to inflation in the landscape of possible vacua; commonly interpreted as an ``effect" of our fundamental ignorance of ST non-perturbative regime. Nevertheless, from BU way, cosmological observables constrain the possible inflation models allowing to distinguish realistic ST vacua. Indeed, a natural possibility is that inflation is driven by one of the axion-like-particles (ALPs) in the String Axiverse \cite{Svrcek:2006yi}.

Axions can generically be originated from dimensional reduction of higher rank p-forms with anomalous couplings. For instance, the axion sector can arise from a 4-form field strength in 11-d SUGRA with Chern-Simons terms after six dimensional compactification \cite{Kaloper:2011jz}. A dimension-5 operator of the axion with a dark $U(1)$ gauge field is also obtained after the dimensional reduction of 11-d SUGRA down to 4-d. 
On the other hand, it is possible that the axion field potential has a periodic monodromic form 
from non-perturbative stringy corrections 
such as world-sheet or Euclidean D-brane instantons \cite{McAllister:2008hb}. Indeed, in ST Axiverse a multi-monodromy inflation scenario can be envisaged. On the other hand, it was shown in many works that secondary Gravitational waves (GWs) and Primordial Black Holes (PBHs) can be efficiently sourced from multi-inflaton dynamics \cite{R1,R2,R3,Espinosa:2018eve,R5,R6,Aldabergenov:2020bpt,Aldabergenov:2020yok}.

D'Amico {\it et al} have shown that the GW spectrum produced in double monodromy inflation can be different than in other scenarios \cite{DAmico:2021vka,DAmico:2021fhz}. In particular, when the axion field is time varying such as during early stage of inflation, one of the gauge field helicity modes becomes tachyonic and exponentially grows until the saturation of the initial inflaton kinetic energy \cite{Anber:2009ua}, sourcing secondary (chiral) tensor modes (GWs). Since this process is followed by second inflationary stage (driven by the second axion), the frequencies of the resulting GWs can fall into the range of the planned space-based GW detectors such as LISA \cite{LISA} and DECIGO \cite{DEC} (in single field and single-stage axion monodromy models, GWs can fall into LIGO frequencies, see e.g. \cite{Sang:2019ndv} and Refs. therein). On the other hand, such a mechanism of gauge field production can modify and amplify the power spectrum of scalar perturbations, leading to additional, scalar-induced GWs and possibly PBHs.

In this paper, we will explore phenomenological implications of the scalar-induced GWs in the double monodromy inflation scenario. By analyzing and comparing the tensor modes sourced by the gauge field instability (which are chiral), and the scalar-induced GWs (which are non-chiral), we show that the double monodromy inflation predicts a characteristic GW signal which is a mix of chiral and non-chiral GWs. We find that the non-chiral contribution can be larger than the chiral one 
under a characteristic frequency threshold which depends on the details of the scalar potential (specifically, on the duration of the first inflationary stage). On the other hand, the chiral GWs rapidly start to dominate above the frequency threshold, and have much larger peak.

\section{Double monodromy model review}\label{sec_2}

\begin{figure}
\centering
  \centering
  \includegraphics[width=.8\linewidth]{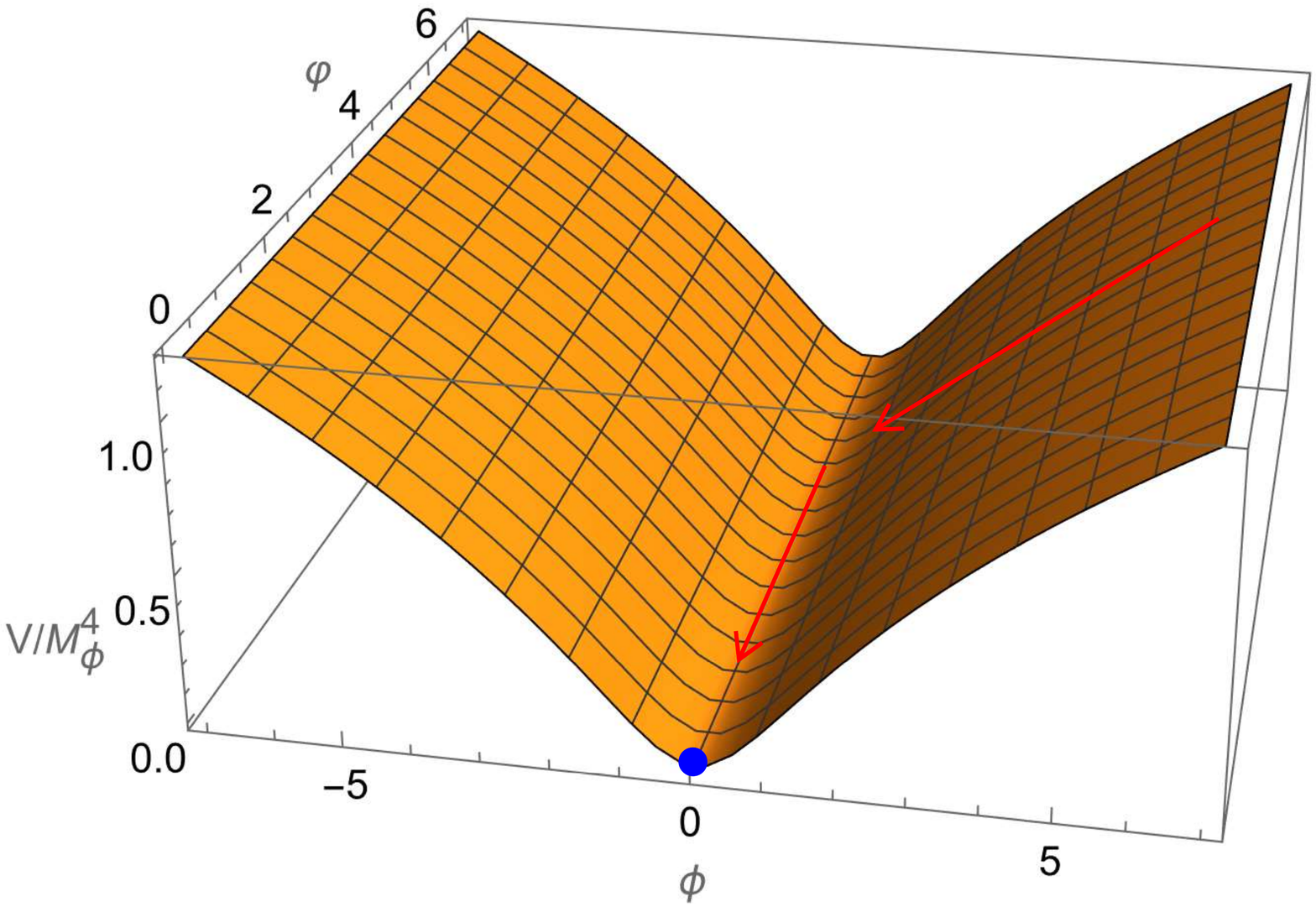}
\captionsetup{width=1\linewidth}
\caption{Scalar potential \eqref{V_full} with the parameter choice $M_\varphi/M_\phi=0.1$, $\mu_\phi=\mu_\varphi=1$, $p=2/5$, and $q=1$. The red arrows show inflationary trajectory during the first and second slow-roll stages, and the blue dot indicates global Minkowski minimum.}\label{Fig_V}
\end{figure}

The double monodromy inflation model proposed in Ref. \cite{DAmico:2021vka} (and further studied in \cite{DAmico:2021fhz}) is based on the Lagrangian (we use ``mostly plus" metric signature and Planck units unless otherwise stated)
\begin{align}
\begin{aligned}
\sqrt{-g}^{\,-1}\cl &=\frac{1}{2}(R-\partial\phi\partial\phi-\partial\varphi\partial\varphi)-\frac{1}{4}F_{mn}F^{mn}\\
&-\frac{\phi}{4f_\phi}F_{mn}\tilde F^{mn}-V(\phi,\varphi)~,
\end{aligned}
\end{align}
where $\phi$ and $\varphi$ are the two axion (pseudo-scalar) fields responsible for two different stages of inflation. $F_{mn}\equiv\partial_mA_n-\partial_nA_m$ is the field strength of the abelian gauge field $A_m$ coupled to the axion $\phi$, with the corresponding axion decay constant $f_\phi$. The dual field strength is given by $\tilde F_{mn}\equiv\tfrac{1}{2}\epsilon_{mnkl}F^{kl}$.~\footnote{\,$\epsilon_{mnkl}$ is the Levi-Civita tensor, related to the Levi-Civita symbol $\varepsilon_{mnkl}$ as $\epsilon_{mnkl}=\sqrt{-g}\varepsilon_{mnkl}$, $\epsilon^{mnkl}=\varepsilon^{mnkl}/\sqrt{-g}$, and we use the convention $\varepsilon^{0123}=-\varepsilon_{0123}=1$.} The scalar potential is inspired by monodromy constructions and given by \cite{DAmico:2021vka,DAmico:2021fhz},
\begin{align}\label{V_full}
\begin{aligned}
    V(\phi,\varphi) &=M_{\phi}^4\left[\left(\phi^2/\mu_{\phi}^2+1\right)^{\frac{p}{2}}-1\right]\\
    &+M_{\varphi}^4\left[\left(\varphi^2/\mu_{\varphi}^2+1\right)^{\frac{q}{2}}-1\right]~,
\end{aligned}
\end{align}
where $M_\phi$, $M_\varphi$, $\mu_\phi$, and $\mu_\varphi$ are mass parameters, and the hierarchy $M_\phi>M_\varphi$ is assumed. In this case the inflationary period can be divided into two slow-roll stages, where during the first stage $\varphi$ is fixed at a constant value and inflation is driven by $\phi$. When $\phi$ approaches zero, the first slow-roll stage ends, and the inflationary trajectory starts moving in $\varphi$-direction towards the global minimum at $\varphi=\phi=0$. This is demonstrated in Figure \ref{Fig_V} where we show the scalar potential \eqref{V_full} with $M_\varphi/M_\phi=0.1$, $\mu_\phi=\mu_\varphi=1$, $p=2/5$, and $q=1$ (see also Figure 1 of \cite{DAmico:2021vka}). The CMB scale perturbations are assumed to be generated during the first ($\phi$-driven) stage when $\varphi$ is fixed and its contribution to the energy density is subdominant (since $M_\phi>M_\varphi$), and therefore isocuravature effects and non-gaussianities are ignored at this stage. Our main focus will be the first slow-roll stage determined by the dynamics of $\phi$, since gravitational waves produced at the end of it can be accessible by the space-based detectors as mentioned in Introduction. We therefore ignore the coupling of $\varphi$ to the gauge field, since during the first slow-roll stage $\varphi$ is fixed. Of course the $\varphi F\tilde F$ term could be important during the second stage, and our results below can in principle be directly applied to it as well, but the frequency of the resulting GWs would be much larger, outside the range of LISA and DECIGO (because they would be produced at a later time during inflation).

It is also worth mentioning that $\varphi$ need not be an axion or have the specific scalar potential as in \eqref{V_full}. We only require that it is destabilized after the first slow-roll period, and its potential is suitable to drive the second stage of inflation in the orthogonal direction to $\phi$.

In FLRW spacetime, $\{-1,a(t),a(t),a(t)\}$, the background equations of motion of this model, during the first stage of inflation, are~\footnote{We define the end of the first stage by the condition $\epsilon\equiv -\dot H/H^2=1$.}
\begin{align}
    \ddot\phi+3H\dot\phi-f^{-1}_{\phi}{\bf E}\cdot{\bf B}+\partial_\phi V &=0~,\label{EOM_phi}\\
    3H^2-\tfrac{1}{2}\dot\phi^2-\rho_{EB}-V &=0~,\label{EOM_H}\\
    {\bf A}''-\nabla^2{\bf A}-f_{\phi}^{-1}\phi'\,\nabla\times{\bf A} &=0~,\label{EOM_A}
\end{align}
where $\bf A$ is the vector potential (in Euclidean space) whose components we denote as $A_a$ with $a=1,2,3$. The electric and magnetic fields are defined as ${\bf E}\equiv -\dot{\bf A}/a$ and ${\bf B}\equiv\nabla\times{\bf A}/a^2$ (in Coulomb gauge $\nabla\cdot{\bf A}=A_0=0$), so that we have
\begin{align}
\begin{aligned}
    F_{mn}F^{mn} &=-2({\bf E}^2-{\bf B}^2)~,\\
    F_{mn}\tilde F^{mn} &=-4{\bf E}\cdot{\bf B}~.
\end{aligned}
\end{align}
The square of a three-vector is its dot product with itself, and $\rho_{EB}\equiv({\bf E}^2+{\bf B}^2)/2$. For the time derivatives we use the notation $\dot\phi\equiv\partial_t\phi$ and $\phi'\equiv\partial_\tau\phi$ where $\tau$ is conformal time, $dt/d\tau=a$. As usual, the Hubble function is $H\equiv\dot a/a$. The values of ${\bf E}\cdot{\bf B}$ and $\rho_{EB}$ entering Eqs. \eqref{EOM_phi} and \eqref{EOM_H} should be understood as avaraged over many universes.

It is convenient to rewrite Eq. \eqref{EOM_A} by using the Fourier transform of ${\bf A}({\bf x},\tau)$,
\begin{align}
\begin{aligned}
    {\bf A}({\bf x},\tau)=& \int\frac{d^3k}{(2\pi)^{3/2}}\sum_{\sigma=\pm}\Big[{\bf e}_\sigma({\bf k})\hat a_{\sigma,{\bf k}} A_\sigma({\bf k},\tau)e^{i{\bf k}\cdot{\bf x}}\\
    &\hspace{1.2cm}+{\bf e}^*_{\sigma}({\bf k})\hat a_{\sigma,{\bf k}}^{\dagger}A^*_{\sigma}({\bf k},\tau)e^{-i{\bf k}\cdot{\bf x}}\Big]~,
\end{aligned}
\end{align}
where $\sigma=\pm$ refers to the two circular polarizations, ${\bf e}_\sigma$ is the polarization tensor, $\hat a$ and $\hat a^\dagger$ are annihilation and creation operators. The polarization tensor obeys ${\bf e}_\sigma\cdot{\bf k}=0$, ${\bf e}^*_\sigma={\bf e}^*_{-\sigma}$, ${\bf e}^*_\sigma\cdot{\bf e}_{\sigma'}=\delta_{\sigma\sigma'}$ and $i{\bf k}\times{\bf e}_\sigma=\sigma k\,{\bf e}_\sigma$. This leads to the equations of motion for the Fourier coefficients $A_\sigma({\bf k},\tau)$,
\begin{equation}\label{A_eq}
    A_\sigma''+(k^2-\sigma f_{\phi}^{-1}k\phi')A_\sigma=0~.
\end{equation}
If $\phi'\neq 0$ and quantity in the parentheses is negative, the corresponding mode of the gauge field becomes tachyonic and can be abundantly produced during inflation \cite{Anber:2009ua}. Without loss of generality we can take $\phi'<0$ during slow roll and $f_\phi>0$, so that the mode $A_-$ becomes tachyonic. For further convenience we introduce an auxiliary function,
\begin{equation}
    \xi\equiv\frac{|\partial_N\phi|}{2f_\phi}=\frac{|\phi'|}{2f_\phi\,aH}~,
\end{equation}
where $N$ is the e-fold time variable (we use the convention where $\dot N=H$, i.e. $N$ grows with time). In the regime $k\ll 2\xi aH$ the negative helicity solution to \eqref{A_eq} can be approximated by \cite{Anber:2009ua},
\begin{equation}
    A_-\simeq \frac{1}{\sqrt{2k}}\left(\frac{k}{2\xi aH}\right)^{1/4}\exp\left(\pi\xi-2\sqrt{\frac{2\xi k}{aH}}\right)~,
\end{equation}
while $A_+$ can be ignored (in \cite{Anber:2009ua} the roles of $A_+$ and $A_-$ are reversed). Using this solution, $\rho_{EB}$ and ${\bf E}\cdot{\bf B}$ take the form,
\begin{align}\label{rho_EB_approx}
\begin{aligned}
    \rho_{EB} &\simeq 1.3\times 10^{-4}e^{2\pi\xi}\frac{H^4}{\xi^3}~,\\
    {\bf E}\cdot{\bf B} &\simeq 2.4\times 10^{-4}e^{2\pi\xi}\frac{H^4}{\xi^4}~.
\end{aligned}
\end{align}

Near the end of the first stage of inflation (driven by $\phi$), this leads to the production of chiral gravitational waves, with their density estimated by \cite{DAmico:2021vka,Domcke:2017fix}
\begin{equation}\label{Omega_GW_c}
    \Omega_{\rm GW,C}\simeq\frac{\Omega_{r,0}H^2}{12\pi^2}\left(1+4.3\times 10^{-7}e^{4\pi\xi}\frac{H^2}{\xi^6}\right)~,
\end{equation}
where $\Omega_{r,0}=8.6\times 10^{-5}$ is the radiation density today. Our e-fold time $N$ can be related to the frequency of the gravitational waves $\nu$ as \cite{DAmico:2021vka}
\begin{equation}
    N=44.9+\log\frac{\nu}{10^{-2}{\rm Hz}}~,
\end{equation}
where we choose the normalization of $N$ where $N=0$ corresponds to the moment of horizon exit of the CMB scale $k=0.002~{\rm Mpc}^{-1}$, and we assume that the total duration of inflation is $60$ e-folds. As usual, the wavenumber-frequency relation is $k=2\pi\nu$ (in natural units). In \cite{DAmico:2021vka} the authors show that the (chiral) GW density estimated by \eqref{Omega_GW_c} peaks at the frequencies around $0.005$ Hz, corresponding to the end of the first stage of inflation (provided it lasts around $30-40$ e-folds) and accessible by the planned LISA experiment \cite{LISA}. In the following section we will plot both the result of \eqref{Omega_GW_c} as well as the total GW density after taking into account secondary gravitational waves induced by the enhanced scalar perturbations.

\section{Scalar-induced (non-chiral) gravitational waves}\label{sec_3}

\begin{figure}
\centering
  \includegraphics[width=.8\linewidth]{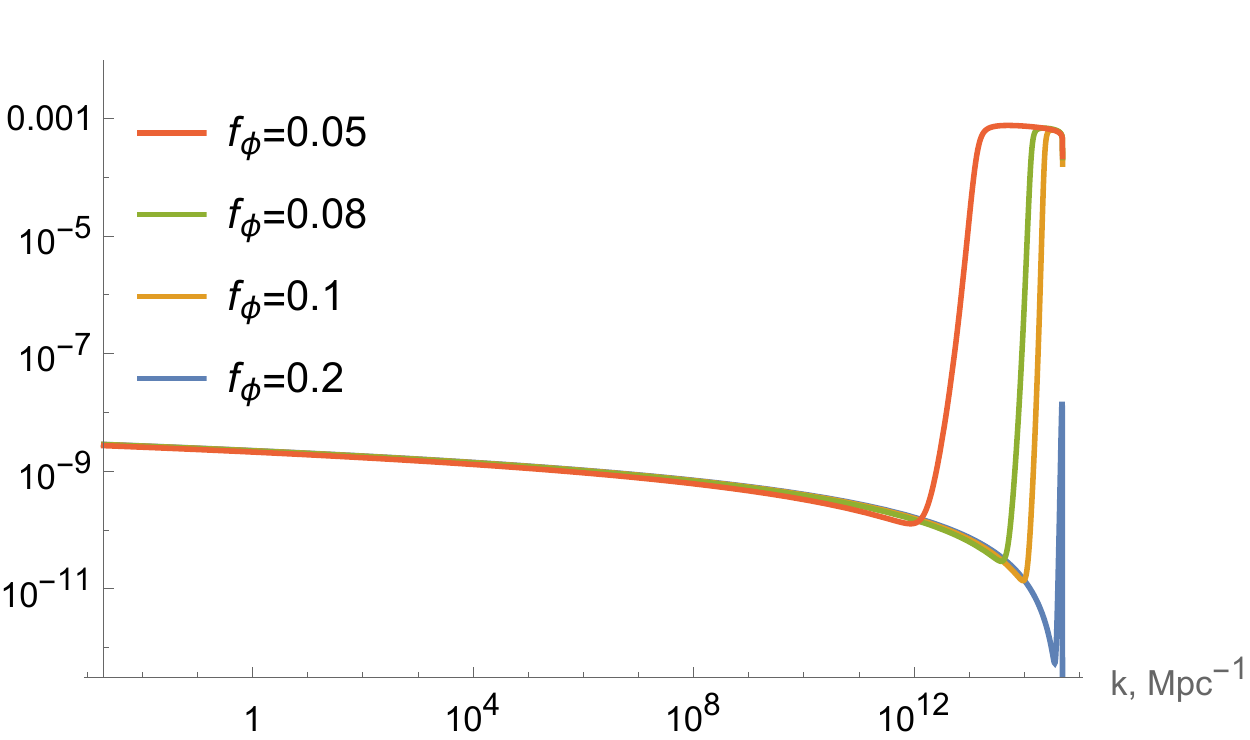}
\captionsetup{width=1\linewidth}
\caption{Power spectrum of scalar perturbations leaving the horizon during the first stage of inflation $\Delta N_2=40$. The parameters are $M_\phi=2\times 10^{-9}$, $\mu_\phi=1$, and $p=2/5$.}\label{Fig_P}
\end{figure}

Taking into account the strong gauge field regime (at the end of the first inflationary stage), the scalar power spectrum can be estimated by \cite{Domcke:2017fix,Domcke:2016bkh,Barnaby:2010vf}
\begin{equation}\label{P_spec}
    P_\zeta\simeq \left(\frac{H^2}{2\pi|\dot\phi|}\right)^2+\left(\frac{{\bf E\cdot B}}{3f_\phi H\dot\phi-2\pi\xi\,{\bf E\cdot B}}\right)^2~,
\end{equation}
where the first term is the standard expression from slow-roll approximation, while the second term is the gauge field contribution. After numerically solving equations of motion \eqref{EOM_phi} and \eqref{EOM_H} by using the approximation \eqref{rho_EB_approx}, we plot the power spectrum \eqref{P_spec} in Figure \ref{Fig_P}, and choose the parameters $M_\phi=2\times 10^{-9}$, $\mu_\phi=1$, and $p=2/5$, which are used in Ref. \cite{DAmico:2021vka}. The duration of the first inflation is set to $\Delta N_1=40$ (we use $\Delta N_1$ and $\Delta N_2$ to denote the lengths of the first and second inflationary stages, respectively).

The peak in the power spectrum is quite large, although our estimates using Press-Schechter formalism \cite{Press:1973iz} show that it is not enough to produce any significant PBH abundance.~\footnote{An efficient PBH production by the similar mechanism (from gauge field instability) was considered in Ref. \cite{Domcke:2017fix} where the necessary power spectrum enhancement was achieved with the help of non-minimal gravitational coupling of the axion.} More specifically, we used the formulae given in \cite{Inomata:2017okj} and references therein to compute the density contrast from a given scalar power spectrum, and a range of critical density values of $\delta_c=0.4\sim 2/3$, all of which produced negative results. Nevertheless, this power spectrum enhancement could be enough to induce potentially detectable stochastic GW background.

The scalar-induced gravitational wave density can be estimated by \cite{Espinosa:2018eve,Kohri:2018awv,Domenech:2019quo,Domenech:2021ztg}
\begin{widetext}
\begin{equation}\label{Omega_GW_nc}
    \Omega_{\rm GW,NC}(k)=\frac{c_g\Omega_{r,0}}{72}\int^{\frac{1}{\sqrt{3}}}_{-\frac{1}{\sqrt{3}}}{\rm d}d\int^{\infty}_{\frac{1}{\sqrt{3}}}{\rm d}s\left[\frac{(s^2-\frac{1}{3})(d^2-\frac{1}{3})}{s^2+d^2}\right]^2 P_\zeta(kx)P_\zeta(ky)\left(I_c^2+I_s^2\right)~,
\end{equation}
\end{widetext}
where $c_g\approx 0.4$ for Standard Model field content \cite{Espinosa:2018eve}, and ${\rm NC}$ stands for ''non-chiral". The variables $x,y$ are related to the integration variables $s,d$ as
\begin{equation}
    x=\tfrac{\sqrt{3}}{2}(s+d)~,~~~y=\tfrac{\sqrt{3}}{2}(s-d)~,
\end{equation}
and the functions $I_c$ and $I_s$ are given by
\begin{align}
\begin{aligned}
    I_c &=-4\int^{\infty}_0{\rm d}\tau\sin{\tau}\big\{ 2T(x\tau)T(x\tau)\\
    &+\big[T(x\tau)+x\tau T'(x\tau)\big]\big[T(y\tau)+y\tau T'(y\tau)\big]\big\} ~,
\end{aligned}\\
\begin{aligned}
    I_s &=4\int^{\infty}_0{\rm d}\tau\cos{\tau}\big\{ 2T(x\tau)T(x\tau)\\
    &+\big[T(x\tau)+x\tau T'(x\tau)\big]\big[T(y\tau)+y\tau T'(y\tau)\big]\big\} ~,
\end{aligned}
\end{align}
where
\begin{equation}
    T(k\tau)=\frac{9}{(k\tau)^2}\left[\frac{\sqrt{3}}{k\tau}\sin\left(\frac{k\tau}{\sqrt{3}}\right)-\cos\left(\frac{k\tau}{\sqrt{3}}\right)\right]~,
\end{equation}
Integrating $I_c$ and $I_s$ yields
\begin{align}
    I_c&=-36\pi\frac{(s^2+d^2-2)^2}{(s^2-d^2)^3}\theta(s-1)~,\\
    I_s&=-36\frac{s^2+d^2-2}{(s^2-d^2)^2}\left[\frac{s^2+d^2-2}{s^2-d^2}\log\left|\frac{d^2-1}{s^2-1}\right|+2\right]~,
\end{align}
where $\theta$ is the Heaviside step function.

\begin{figure*}
\centering
  \centering
  \includegraphics[width=1\linewidth]{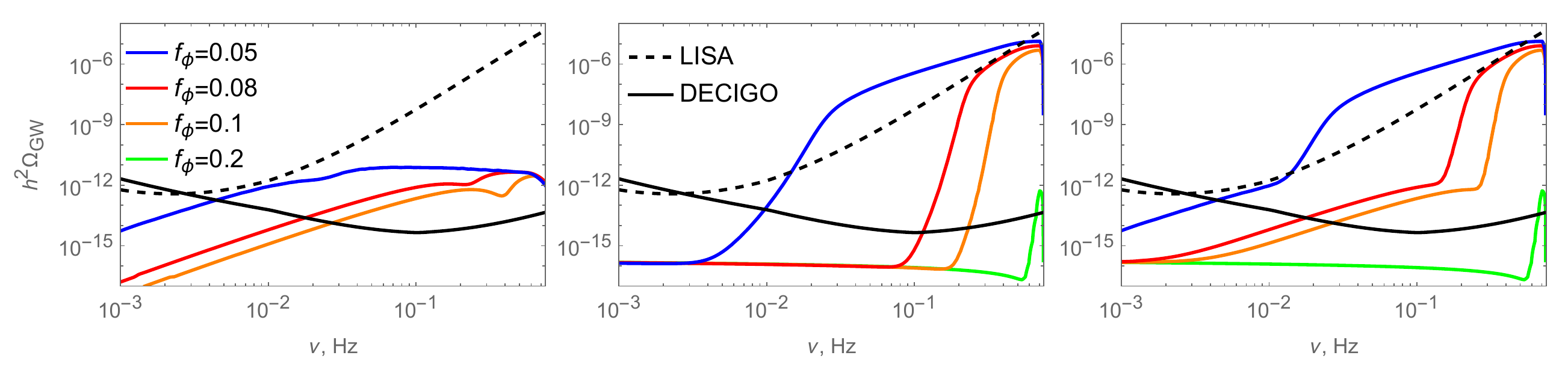}
\captionsetup{width=1\linewidth}
\caption{Left: Scalar-induced GW density calculated by Eq. \eqref{Omega_GW_nc} (for $f_\phi=0.2$ the GW density is negligible and cannot be seen in this plot). Center: Chiral GW density from Eq. \eqref{Omega_GW_c} (sourced by the gauge field). Right: Total GW density \eqref{Omega_tot} with both chiral and non-chiral contributions. Solid and dashed black curves represent DECIGO and LISA expected sensitivities, respectively.}\label{Fig_GW}
\end{figure*}

Substituting the power spectrum \eqref{P_spec} into the induced GW density \eqref{Omega_GW_nc}, we can estimate the non-chiral GW abundance induced by scalar perturbations -- see Figure \ref{Fig_GW} (left) (the parameters are the same as in Figure \ref{Fig_P}). We plot the expected LISA and DECIGO sensitivities as dashed and solid black curves, respectively. It can be seen that for $\Delta N_1=40$ the GW density peaks are within the DECIGO sensitivity. If we choose smaller $\Delta N_1$, say $\Delta N_1=35$, the peaks will also enter LISA's frequencies of around $10^{-3}$ to $10^{-2}$ Hz, although a smaller range of these GWs would be detectable due to the lower sensitivity of LISA.

In Figure \ref{Fig_GW} (center) we show the chiral GW contribution \eqref{Omega_GW_nc}, and Figure \ref{Fig_GW} (right) shows the total GW density estimated as the sum of chiral and non-chiral contributions,
\begin{equation}\label{Omega_tot}
    \Omega_{\rm GW,total}=\Omega_{\rm GW,C}+\Omega_{\rm GW,NC}~,
\end{equation}
As can be seen, the higher frequencies of total GW spectrum are dominated by the chiral GWs which drop off quickly at lower frequencies, where the scalar-induced, non-chiral GWs provide the dominant contribution. To be specific, let us take an example of $f_\phi=0.1$ (orange curves). In this case, the frequencies of around $1$ Hz are mostly chiral GWs whose density parameter is reaching close to $10^{-5}$. At around $0.1$ Hz, the chiral GW contribution drops to $\sim 10^{-16}$, and the main contribution comes from the induced GWs. Although the density parameter for the latter reaches only $\sim 10^{-13}$, it is still within the expected DECIGO sensitivity. We therefore have an interesting possibility that different (but neighbouring) frequencies of the potentially detectable GW background can have different polarizations. This could be seen as a distinguishing signature of this class of models, in comparison to theories where detectable gravitational waves are purely/mostly chiral or non-chiral. Another model leading to observable GWs as a combination of different chiralities (but with qualitatively different spectrum) was proposed in \cite{Odintsov:2021kup} based on modified gravity.

\section{Conclusions and remarks}

In this letter we have studied in more detail the phenomenology of gravitational waves of the double axion monodromy inflation proposed in \citep{DAmico:2021vka}. We showed that in addition to chiral GWs sourced by the gauge field, the model also predicts GWs produced by scalar perturbations which are in turn enhanced by the gauge field instability. These scalar-induced GWs are non-chiral, and peak at around the same frequency as the chiral modes as shown in Figure \ref{Fig_GW}. And although the density of scalar-induced GWs is much smaller at its peak, it decreases much more slowly than the density of chiral GWs, as we decrease the frequency. This allows the non-chiral modes to dominate the total tensor spectrum up to a characteristic frequency threshold (depending on the duration of the first stage of inflation) at which point the chiral GW density experiences a sudden increase and overtakes the total GW density. The combination of the two contributions renders the GW signal visibly different than in other double inflation scenarios, which could be tested by the space-based interferometers such as LISA and DECIGO in the (hopefully) near future. In this sense, GW physics can be crucial for the Axiverse exploration within the String Landscape.

Regarding possible generalizations of the model, the scalar field driving the second inflationary stage (which we call $\varphi$) may or may not be an axion, since the only requirement is that its scalar potential has a suitable shape for inflation. It may also be interesting to consider non-minimal couplings of the scalars to gravity along the lines of Ref. \cite{Domcke:2017fix}, or with each other, which can further enhance the scalar perturbations and scalar-induced gravitational waves, as well as open up the possibility of (asteroid-mass) primordial black holes as dark matter.

\begin{acknowledgments}
The work by M.A., A.Ab., Y.A., and D.B. was supported by the Science Committee of the Ministry of Education and Science of the Republic of Kazakhstan (Grant \# BR10965191 ``Complex research in nuclear and radiation physics, high-energy physics and cosmology for development of the competitive technologies"). 
A.Ad. work is supported by the Talent Scientific Research Program of College of Physics, Sichuan University, Grant No.1082204112427 \& the Fostering Program in Disciplines Possessing Novel Features for Natural Science of Sichuan University, Grant No.2020SCUNL209 \& 1000 Talent program of Sichuan province 2021.
Y.A. was supported in part by Thailand NSRF via PMU-B [grant number B05F650021] and Thailand Science research and Innovation Fund Chulalongkorn University CU$\_$FRB65$\_$ind (2)$\_$107$\_$23$\_$37.
\end{acknowledgments}

\end{document}